\documentclass[runningheads]{llncs}

\usepackage[T1]{fontenc}
\usepackage{graphicx}

\usepackage[margin=1in]{geometry}

\usepackage[utf8]{inputenc}
\usepackage{stmaryrd }
\usepackage{hyperref}
\usepackage[english]{babel}
\usepackage{blindtext}
\usepackage{cite}
\usepackage{amsmath,amssymb,amsfonts,stackengine}
\usepackage{tikz}
\usetikzlibrary{shapes,fit,shapes.geometric, angles}

\usepackage{circuitikz}
\usepackage{algorithmic}
\usepackage{graphicx}
\usepackage{textcomp}
\usepackage{mathtools}
\usepackage{xcolor}
\usepackage{multirow}
\usepackage{lipsum}
\usepackage{amsthm}

\stackMath
\usepackage{ifthen}

\newcounter{sqindex}

\newcommand{\hexagon}{\mathord{\raisebox{0.6pt}{\tikz{\node[draw,scale=.65,regular polygon, regular polygon sides=6,fill=none](){};}}}}

\newcommand\leavesof[1]{\texttt{LEAF}_{#1}}

\begin{document}
\title{Towards Interdependent Safety Security Assessments using Bowties}

\author{Luca Arnaboldi \and
David Aspinall }
\institute{School of Informatics, University of Edinburgh, Edinburgh, UK \\
\email{\{luca.arnaboldi,david.aspinall\}@ed.ac.uk}}
\authorrunning{L. Arnaboldi and D. Aspinall}

\maketitle
\begin{abstract}
We present a way to combine security and safety assessments using Bowtie Diagrams.
Bowties model both the \emph{causes} leading up to a central failure event %
and \emph{consequences} which arise from that event, as well as \emph{barriers} which
impede events.   
Bowties have previously been used separately for security and safety assessments, but we
suggest that a unified treatment in a single model can elegantly capture safety-security
interdependencies of several kinds.
We showcase our approach with the example of the October~2021 Facebook~DNS shutdown, examining the chains of events and the interplay between the security and safety barriers which caused the outage.

\end{abstract}

\newcommand{\nodetype}[1]{\ensuremath{\mathtt{#1}}\xspace}
\newcommand{\LEAFt}{\nodetype{LEAF}}
\newcommand{\ANDt}{\nodetype{AND}}
\newcommand{\ORt}{\nodetype{OR}}
\newcommand{\INHIBITt}{\nodetype{INHIBIT}}
\newcommand{\CHOOSEt}{\nodetype{CHOOSE}}
\newcommand{\theset}[1]{\{\,#1\,\}}
\newcommand{\ch}{\mathop{ch}}
\newcommand\choicesof[1]{\ensuremath{\mathtt{CHOOSE}_{#1}}}
\newcommand{\TRUE}{\ensuremath{\mathbf{1}}}
\newcommand{\FALSE}{\ensuremath{\mathbf{0}}}
\newcommand\addvmargin[1]{
  \node[fit=(current bounding box),inner ysep=#1,inner xsep=0]{};
}

\section{Introduction}

Structured graphical reasoning methods help the systematic discovery, assessment and mitigation of risks.
A variety of methods have been used for security and for safety.
While safety usually considers unintentional events and security deals with intentional malicious events,
both safety and security assessments may consider causes and consequences of critical events.
An early method is cause/consequence diagrams~\cite{nielsen1971cause}, commonly
referred to as the ``bowtie method'' due to its graphical representation.
A bowtie is formed of a tree of
causes and a tree of consequences, fanning out to the left and right of a central critical (or \emph{top}) event.
Nodes in the tree represent causal or following events and \emph{barriers} represent ways
to prevent or mitigate events.
Bowties
have been used both for safety~\cite{shahriar2012risk} and security~\cite{bernsmed2017visualizing} separately, and an initial investigation was even performed to inform a safety bowtie by an attack tree~\cite{abdo2018safety} in terms of risk quantification.
They %
have many desirable properties, such as ease of visualisation~\cite{bernsmed2017visualizing} and the
ability to quantify and calculate likelihoods of events, given a mathematical definition for a diagram's meaning.

In this paper we introduce a way to combine safety and security assessments together in a single
bowtie, to enable reasoning about security and safety together.  In particular we account for four
different types of interplay between event types: safety and security can act independently, with
conditional dependency, in concert, or in opposition ~\cite{kriaa2015survey}.  An ideal structured
reasoning approach should capture each of these.

\paragraph{2021 Facebook DNS outage.} We use Facebook's recent global DNS outage as a running example,
which nicely demonstrates some safety-security interplay~\cite{janardhan2021first,janardhan2021second}. 
On October 4$^{\text{th}}$ 2021, there was a complete outage of all Facebook apps and services (including WhatsApp, messenger, Instagram, etc.), which lasted approximately 6 hours. 
An initial small change to the DNS configuration triggered a complicated chain of events that made resolving a seemingly small issue overly complex and time consuming, and affected millions of people. %
This happened directly as an outcome of a security response disabling the ability for the safety engineers to patch the server remotely, forcing them to physically access the servers, which exacerbated the disruption.

\begin{figure} [htbp!]
    \centering
    \includegraphics[width=\textwidth]{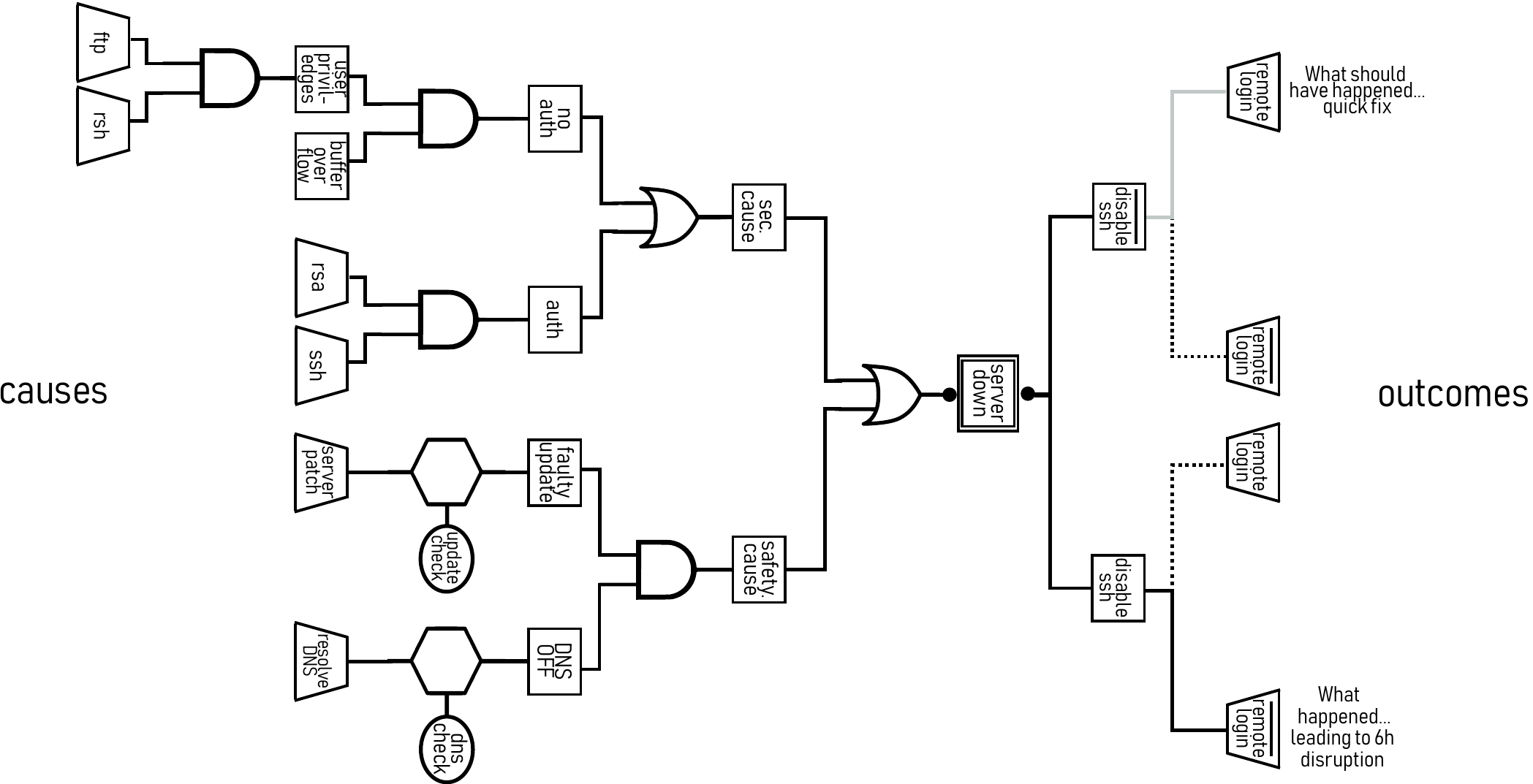}
    \caption{High level bowtie diagram of the Facebook DNS outage (Tab.~\ref{tab:edge-labels} expands the labels). The negative outcome is in black, whilst the grey path highlights the path that would avoid the negative outcome. Showing the antagonism between the security and safety responses.}
    \label{fig:antagonistic}
\end{figure}

A bowtie describing the sequence of events is presented in Fig.~\ref{fig:antagonistic}.
In outline it shows that the DNS server-down event (the top event) could have a security cause
or a safety cause.  For example, a safety cause could be that a faulty update to the server
configuration goes wrong; a barrier to stop this would be to check the patch doesn't cause
the server to fail.  A security cause might be that an attacker manages to access the DNS
server and make a malicious update.  If the DNS server does go down, we would like to find a path to a
rapid fix to make the server operational again.  But two fixes, a safety
fix and a security fix, are \emph{antagonistic} and act in opposition; only one of them
should be undertaken.  In the Facebook DNS incident, both fixes were implemented,
disallowing remote access to the servers and delaying the response.

Obviously this bowtie is not a complete safety-security
assessment (and was constructed after-the-fact).  A more complete model
could cover the many bad consequences that can occur from a central
DNS outage and ways those could be mitigated, as well as further ways
such outages could occur.  But this example serves to demonstrate
how we can model some safety-security interdependences and we will give
mathematical definitions which formalise this, returning to the example
and variations of it in Sec.~\ref{sec:dependencies}.

\paragraph{Outline.} The rest of the paper is structured as follows: Sec.~\ref{sec:background} contains background on diagrammatic representation of bowties, we note that further background is added as needed to the respective sections, Sec.~\ref{sec:formalisation} the formalisations of safety-security bowties, Sec.~\ref{sec:dependencies} the interdependence analysis and the formalisation of its operations,  and in Sec.~\ref{sec:conclusions} we provide examples of related work, related assessment techniques, and discuss next steps for this work.

\paragraph{Contributions.} The contributions of this work are the following:
\begin{enumerate}
\item A demonstration of security and safety interplay using bowtie diagrams and
  the Facebook DNS outage.  %
  Fig.~\ref{fig:antagonistic} highlights the need to treat safety and security together, 
  before a critical event and afterwards.
\item Expanding on recent work on combining and formalising safety-security trees as
  \emph{disruption trees}~\cite{stoelinga2021marriage}, we add barriers (safety) and
  mitigations (security), modelling both response forms in  the same formalism.
\item We also add a corresponding mechanism for outcomes, to give a
  new definition for safety-security bowtie diagrams.  %
  This provides a solid underlying basis for the graphical notation,
  paving the way for further theoretical analysis and discussion, and
  ultimately towards trustworthy tools to reason about and manipulate such diagrams.
\end{enumerate}
This paper is a first step and we want to investigate more complex
case studies as well as studying properties of the definitions,
see Sec.~\ref{sec:conclusions} for more discussion of next steps.

\section{Diagrammatic Representations}
\label{sec:background}

\renewcommand{\arraystretch}{1.2}
\begin{table}[htb!]
    \centering
        \caption{Symbols and Relationships in a fault tree}
    \begin{tabular}{c|p{0.7\columnwidth}}
    \hline
    \multicolumn{2}{c}{\textbf{Event Types}}\\\hline
     \noalign{\smallskip}
    \centering   \begin{tikzpicture}[baseline=0, scale=.4] \draw (0,0) rectangle (2,1); \end{tikzpicture} &   \textit{Intermediate event} - an identity function labelling outgoing edges. \\
    
    \centering   \begin{tikzpicture}[baseline=0] \draw (.2,0.2) ellipse (10pt and 6pt); \end{tikzpicture} {}  &   \textit{Prevention } - restrictive conditions, must fail to progress \vspace{0em}\\
    
     \centering   \begin{tikzpicture} \node [trapezium, trapezium angle=60, minimum width=25, draw]  {};\end{tikzpicture} &   \textit{LEAF} - initiation event (BEs in Fault Trees) eventual outcome (in Event Trees) 
    \\     \hline
   \multicolumn{2}{c}{\textbf{Relationship Types}}\\\hline
          \noalign{\smallskip}
            \begin{minipage}{.06\textwidth}
                     \includegraphics[width=\textwidth]{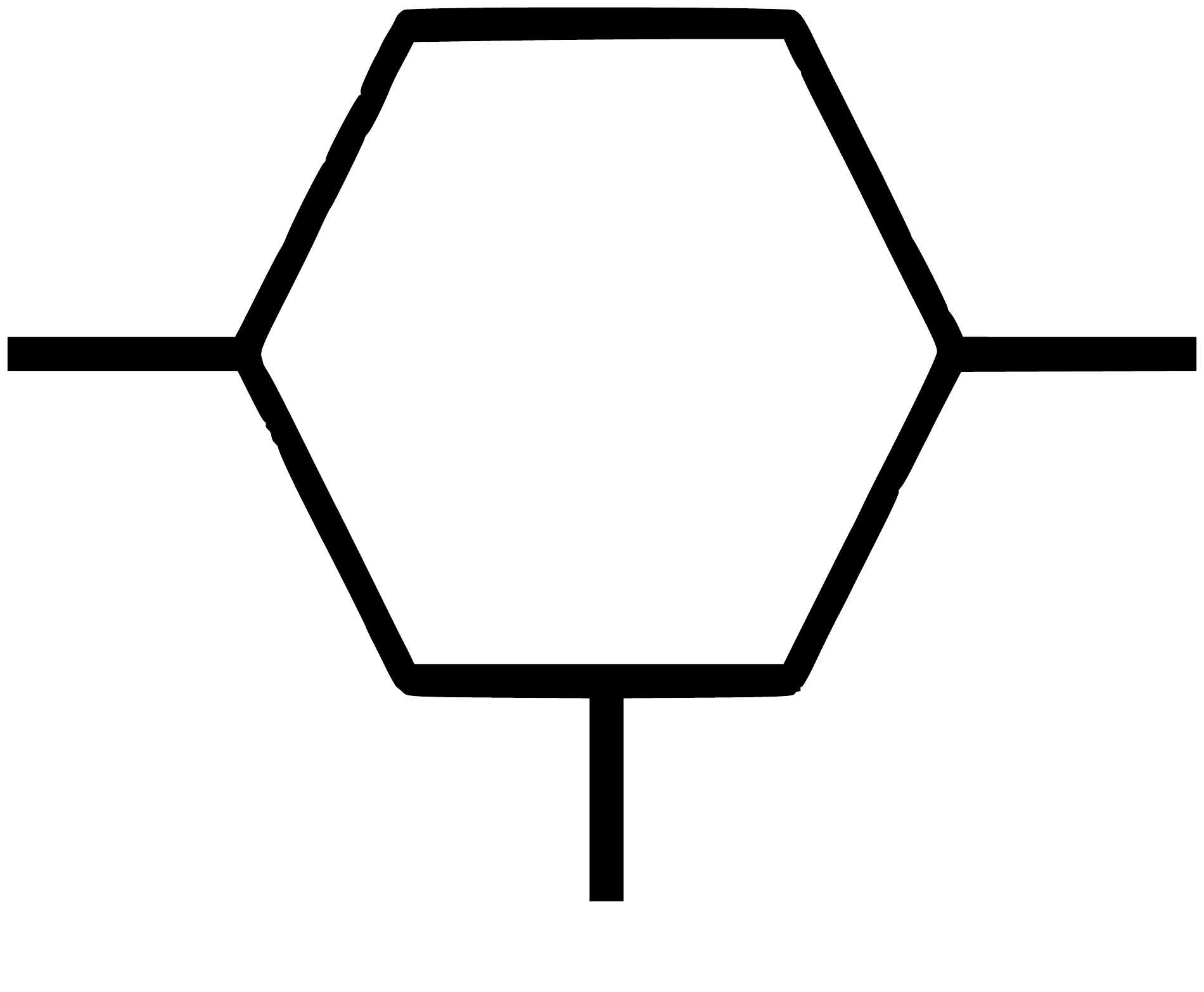}
            \end{minipage}

        & INHIBIT - output occurs if the LH occurs and prevention does not \\ %
    \resizebox{35pt}{15pt}{%
   \begin{circuitikz} \draw
        (2,3) node[or port, fill=none] (myor) {}
        
        (myor.in 1) node[left =.2cm] (a) { }
        (myor.in 2) node[left =.2cm] (b) { }
        (myor.out)  node[right=.2cm] (c) { }
        
        (myor.in 1) -- (a)
        (myor.in 2) -- (b)
        (myor.out)  -- (c);
        \end{circuitikz}}&   OR -  output occurs if at least one of the inputs occurs\\
        \resizebox{35pt}{15pt}{%
   \begin{circuitikz} \draw
        (1.2,2.2) node[and port, fill=none] (myor) {}
        
        (myor.in 1) node[left =.2cm] (a) {}
        (myor.in 2) node[left =.2cm] (b) {}
        (myor.out)  node[right=.2cm] (c) {}
        
        (myor.in 1) -- (a)
        (myor.in 2) -- (b)
        (myor.out)  -- (c);
        \end{circuitikz}} &   AND - output occurs if both of the inputs occur\\

 \hline
    \end{tabular}

    \label{tab:symbols}
  \end{table}

Our bowtie diagrams use Boolean logic gate notation to show relationships between events tracking the passage of failures from ``inputs'' to ``outputs''.
Boxes containing labels describe intermediate events. 
\paragraph{Faults and attacks in the same diagram.} 
Security analysis
traditionally uses attack trees~\cite{schneier1999attack} whilst Safety analysis makes use of fault trees~\cite{vesely1981fault}. One shows the steps
which an attacker can take to achieve a goal whilst the other shows the events leading to a fault.  
Stoelinga, however, has
recently argued that fault trees and attack trees can be treated uniformly in a single structure~\cite{stoelinga2021marriage}.
Tab.~\ref{tab:symbols} shows the symbols we use, following the Fault
Tree Handbook notation~\cite{vesely1981fault}.  
\paragraph{Barriers and defences treated uniformly.}
We further propose to also treat barriers and defences in the same logic.  To
capture actions or conditions that impede fault conditions arising or
defences in an attack tree, we use an \texttt{INHIBIT} gate drawn as a
hexagon.  The orthogonal side input is the inhibiting condition which
prevents the flow of the fault or prevents the attack.

\paragraph{Safety Security Bowties.}
Now that we have a unified tree notation for attack trees and fault trees, we can use event tree tree notation for consequences as usual.  This stays true to the essence of bowties as previously
defined~\cite{de2016bowtie}.
The focal point of the assessment is the critical risk event (or ``top event'', following fault tree parlance).
To the left of the event are potential causes of the risk and ways to reduce them; this is the \emph{prevention assessment}.
On the right on the other hand we observe the outcomes of the risk event known as the \emph{consequence assessment}.

\begin{figure}[htb]
     \centering
     \includegraphics[width=0.7\linewidth]{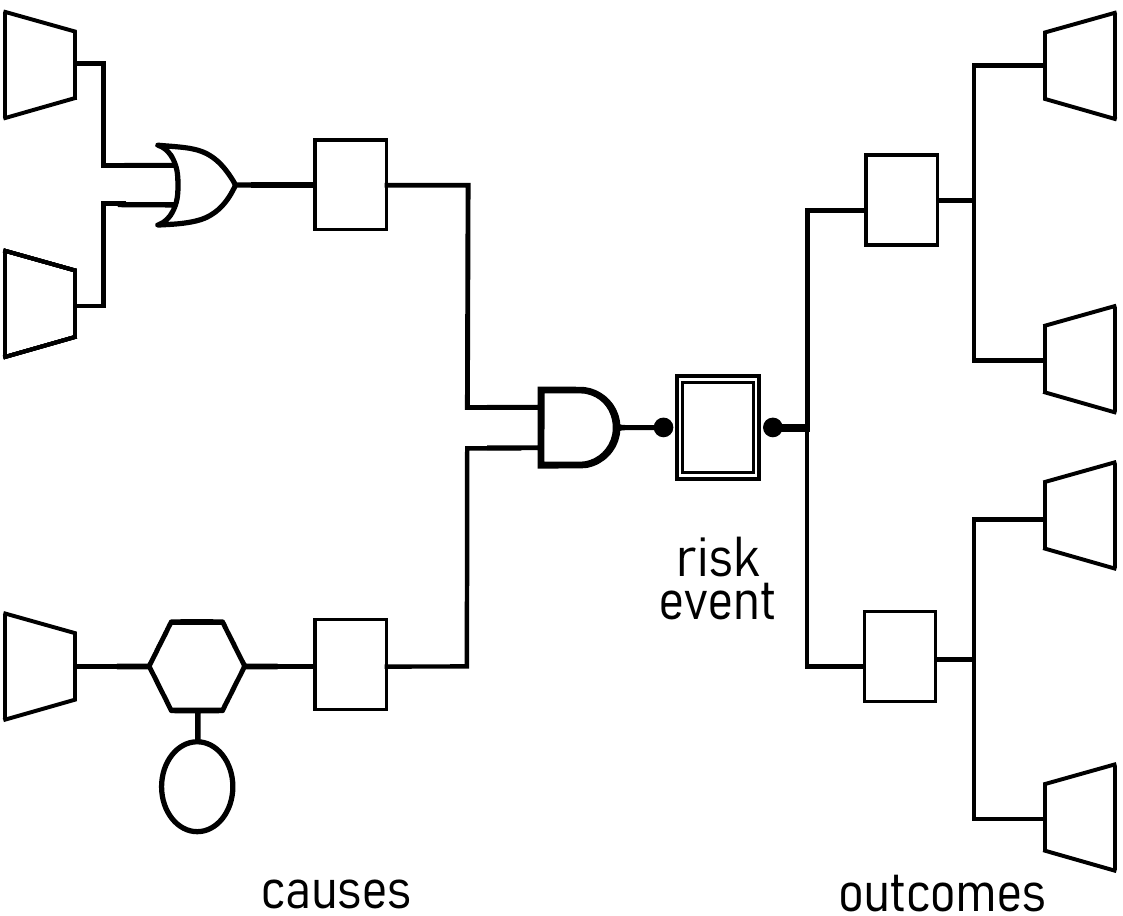}
     \caption{Generic representation of Bowtie Diagram with causes on the left and outcomes on the right.}
     \label{fig:bowtie-genera}
 \end{figure}

An example unlabelled bowtie diagram is shown in Fig.~\ref{fig:bowtie-genera}.
These diagrams are a Boolean network which are static and capture only logical dependencies: a bowtie itself is not quantitative,
but it is a qualitative model that may be analysed quantitatively by adding additional annotations representing risk assessments.  This has been done by various authors for fault and attack trees, 
for example~\cite{ren2017fault,fila2020exploiting}.  
For simplicity in the rest of this paper, we consider the basic logical interpretation only.

\section{Formalisation of Safety-Security Bowties}
\label{sec:formalisation}

Like typical bowties a safety-security bowtie will be composed of a \textit{prevention assessment} and a \textit{consequence assessment}.
For the prevention assessment we use a \emph{disruption prevention tree} which captures safety and security assessments, as well as 
attack and fault mitigations, which correspond to barriers in traditional bowties~\cite{de2016bowtie}.
For the consequence assessment we use an event tree to show the consequences of the risk event and potential outcomes. 
The graphical representation from the figures includes intermediate events between gates, for enhanced understanding, but these are not part of the formalisation.
We first showcase a general structure for trees we make use of for all subsequent formalisations in Sec.~\ref{sec:structure-tree}.
We note that although, we are mostly concerned with trees, we adopt Directed Acyclic Graphs (DAGs) to allow input and outputs to be shared between gates.

\subsection{Structure trees}
\label{sec:structure-tree}
Our models are based on generalised n-ary trees with additional
structure given by a node type labeling. 

\begin{definition}[Structure Tree]
\label{def:st}
A Structure Tree is a tuple $T = \langle N,\mathbb{T},t,\ch\rangle$ where:
\begin{itemize}
    \item $N$ is a finite set of nodes and $\mathbb{T}$ is a set of node types;
    \item $t: N \rightarrow \mathbb{T}$ gives the type of each node;
    \item $\ch: N \rightarrow N^*$ gives the sequence of children of a node.
\end{itemize}
Moreover, T satisfies the following constraints:
\begin{itemize}
    \item $(N,E)$ is a connected DAG where $E = \theset{(v,u) \in N^2 \;|\; u\in \ch(v)}$
    \item $T$ has a unique root, denoted by $R_T:\exists ! R_T \in N . \forall v \in N. R_T \not\in\ch(v)$ (where $\exists !$ stands for unique existence).

\end{itemize}
\end{definition}

Given a type $\LEAFt\in\mathbb{T}$, we write $\leavesof{T} = \theset{v\in N \;|\; t(v)=\LEAFt}$
for the nodes labelled with that type, and similarly for other types.

\subsection{Disruption Trees}

The notion of Disruption Trees (DTs) is introduced by Stoelinga as the
uniform representation of static safety and security risk models~\cite{stoelinga2021marriage,budde2021attack}.
Although extensions exists to each of these assessments which do result in key differences, in the simplest setting with only \texttt{OR} and \texttt{AND} nodes, they coincide.\footnote{There are subtleties in different
  ways to treat \texttt{OR} gates with quantified safety or security risks, which will not be considered here.}
Following Def.~\ref{def:st} we can interpret Stoelinga's DTs as:

\begin{definition}[Disruption Tree \cite{stoelinga2021marriage}]
  \label{def:dt2}
  A Disruption Tree is a structure tree according to Def.~\ref{def:st} with %
  $\mathbb{T} = \theset{\LEAFt,\ANDt,\ORt}$
  and such that $t(v)=\LEAFt$ if{}f $v\in\leavesof{T}$.
\end{definition}

Semantics pin down the mathematical meaning of models. The basic semantics of a disruption tree is given
by its \emph{structure function} over Booleans $\mathbb{B} = \{1, 0\}$, which defines a function
from leaf events to the root in the obvious way.  
Structure Trees are more lax compared to models we usually want.
 Besides allowing useful DAG structure they include redundancies like
nodes with repeated children.  This is alright for \texttt{AND} and \texttt{OR} but
 would be confusing later on with gates that have some inputs (or
 outputs) being negated (or mutually exclusive).  
We call a
 structure tree (and later variants) \emph{proper} if every child of a
 node is distinct, i.e.,
 $|\theset{\ch(v)\;|\;v\in N}|=\mathit{length}(\ch(v))$.
 From now on we will restrict to models built on proper structure trees.

\subsection{Prevention Assessments}
\label{sec:left}

We turn a Disruption Tree
into a Disruption Prevention Tree by adding a new gate type for \texttt{INHIBIT}.

\begin{definition}[Disruption Prevention Tree]
  \label{def:def-dpt}
  A Disruption Prevention Tree (DPT) is a Disruption Tree
  with an additional binary node type
  $\INHIBITt$, i.e.,
  $\mathbb{T} = \theset{\LEAFt,\ANDt,\ORt,\INHIBITt}$ and
  $|\ch(v)|=2$ whenever $t(v)=\INHIBITt$.  The second child of
  an \INHIBITt node corresponds to
  the inhibiting (prevention) condition.
\end{definition}  

There is a similarity between Disruption Prevention Trees and Attack-Defence Trees. 
Formalisations of Attack-Defence Trees (see~\cite{kordy2010foundations} for example) label nodes to represent their role in the model as a proponent
(mitigating an attack or protecting against a fault event) or an opponent (actors causing damage leading to the root fault).
It seems possible %
to recover this labelling from the root of DPT flipping the role when consider the prevention
input of \texttt{INHIBIT} gates.  %

\subsubsection{Structure of a DPT}

The semantics of a DPT allows a proponent to block an opponent.  These kind of barriers are called ``avoid barriers'' in fault trees~\cite{de2006aramis,guldenmund2006development} and ``defences'' in attack trees~\cite{kordy2010foundations}, but function in the same manner.
Following from the DT semantics~\cite{stoelinga2021marriage}, we generalise the structure function to interpret the
additional gate.

\begin{definition}[Structure function for DPTs] %
  \label{def:DPT-sems}
  The structure function $f_T:N\times2^{\leavesof{T}}\rightarrow\mathbb{B}$ of a DPT $T$ and
  $A \subseteq \text{\texttt{LEAF}}$ is defined by:\\
  \begin{displaymath}
    f_T(v,A) = %
    \begin{cases}
      \TRUE &  \begin{cases}%
                        \mathrm{if}\ t(v)=\ORt & \mathrm{and}\ 
                                    \exists u \in \ch(v).\ f_T(u,A) = \TRUE
                        \\%
                        \mathrm{if}\ t(v)=\ANDt & \mathrm{and}\ 
                      \forall u \in \ch(v). \ f_T(u,A) = \TRUE
                        \\%
                        \mathrm{if}\ t(v)=\INHIBITt & \mathrm{and}\ 
                        f_T(u_1,A) = \TRUE, f_T(u_2,A) = \FALSE
                        \ \mathrm{where}\ \ch(v)=[u_1,u_2] \\
                        \mathrm{if}\ t(v)=\LEAFt & \mathrm{and}\ 
                        v\in A 
               \end{cases}  \\%
      \FALSE & otherwise%
    \end{cases}
  \end{displaymath}
  The interpretation for the overall tree is $f_T(A) = f_T(R_T,A)$.
\end{definition}  
\noindent

We will sometimes use syntactic notation to construct DPTs using \texttt{AND} (written as $\cap$), \texttt{OR} ($\cup$) and \texttt{INHIBIT} ($\hexagon$) as term constructors and label
names to stand for nodes in the graph.  This can be made formal by defining interpretation functions on terms (for example,
see~\cite{jhawar2015attack}).

\subsection{Consequence Assessments}
\label{sec:right}

A consequence assessment constitutes the right hand side of a safety-security bowtie and investigates potential ways to mitigate the central risk event or negative outcomes.
As usually done with bowties, we make use of event trees to investigate each possible outcome of the central risk event.
For the RHS of the bowtie, the leaf nodes in the DCT are potential
final consequence events and each \CHOOSEt node represents a splitting
point where a sub-event can have different outcomes (typically for a
binary choice point, something happens or does not happen).  This is
captured by a \emph{consequence function} $C$ which selects the
outcome event at each \CHOOSEt node, tracing a path through to a final
outcome.  For the Boolean interpretation only a single choice can
happen; for more general interpretations we might assign a possibility
(Y/N/M) against each branch, probabilities or probability
distributions, etc.

\begin{definition}[Disruption Consequence Tree]
  \label{def:dct2}
  A Disruption Consequence Tree (DCT) is a structure tree with gate types $\mathbb{T} = \theset{\LEAFt, \CHOOSEt}$.
\end{definition}

    \begin{definition}[Structure function for DCTs] %
      \label{def:DPC2-sems}
      The structure function $f_T:N\times({\choicesof{T}\to\mathbb{N}})\rightarrow N$ of a DCT $T$ is
      defined by:
      \begin{displaymath}
        f_T(v,C) = 
        \begin{cases}
          v & t(v)=\LEAFt \\
          f_T(u_n,C) & \mathrm{where}\ C(v)=n\ \mathrm{and}\ \ch(v)=[u_1,\ldots,u_n]
        \end{cases}                 
      \end{displaymath}
      where $C$ is a consequence function which selects
      an outcome for each \CHOOSEt node, i.e.,
      $1\leq C(v)\leq \mathit{length}(\ch(v))$ for $v\in\choicesof{T}$.
      The interpretation for the overall tree is $f_T(C) = f_T(R_T,C)$.
    \end{definition}

\subsection{Bowties}

A bowtie is just a pair of two structured trees.  Intuitively, the top event corresponds
to the two roots linked together, a realised disruption.  %
\begin{definition}[Disruption Bowtie]
  A Disruption Bowtie (DB) is a pair $\langle T_P, T_c \rangle$ of a disruption prevention
  tree and a disruption consequence tree.
\end{definition}

\section{Interdependence Analysis}
\label{sec:dependencies}

When assessing the interactions between security and safety we categorize them following the existing literature~\cite{kriaa2015survey,eames1999integration}, they are:
\begin{enumerate}
    \item \textit{conditional dependency}, meaning the safety of the system depends on its security, and conversely, 
    \item \textit{mutual reinforcement}, fulfilment of safety requirements or safety measures contributes to security, and conversely, 
    \item \textit{antagonistic}, when considered jointly, safety and security requirements or measures lead to conflicting situations, and 
    \item \textit{independence}, when the interactions are mutually exclusive or not in interference.  
\end{enumerate}

The interactions, obviously, have a great difference on the result of the assessment and in our case study we see dramatic differences in the same exact events inter-playing in different ways.
We showcase in the scenarios the need for specific operations and describe their semantics in this section. 
We omit the right hand side of the bowtie for those interdependencies which do not strictly require it (\textit{independence} and \textit{conditional}).
Whilst we find that both \textit{reinforcing} and \textit{antagonistic} require the recovery side of the bowtie.
We note that each interdependency condition leads to trees joined in the usual fashion when combining to form a new tree.

\subsection{Running Example: Safety Security Interplay}
\label{sec:example}

As main point of reference we use the reported chain of events by Facebook engineering~\cite{janardhan2021first,janardhan2021second} as well as the external assessment conducted by CloudFlare~\cite{martinho2021}.
We conducted a structured risk assessment of the Facebook DNS shutdown using security safety bowties focusing our analysis on how safety and security measures interacted to lead to the specific events.
Our methodology of analysis was the following: i) we downloaded the analysis of the events released by the Facebook engineering team as well as an external examination by cloud hosting provider CloudFlare which witnessed the events externally. ii) we systematically labelled each brief to extract events, its causes and outcomes and the recovery/preventions that were in place (and failed in this case).
For the sake of case study we choose to adapt an attack tree that reached the same risk event of the server outage, for this we borrow an existing example from~\cite{jhawar2015attack}.
For the sake of diagrammatic explainability edges are labelled by an identity function, which creates the intermediate events, a full table containing all edge labels is presented in Tab.~\ref{tab:edge-labels}.

\begin{table}[htb]
\caption{Explanations of edge and event labels.}
\label{tab:edge-labels}
\centering
\footnotesize
\begin{tabular}{p{0.145\linewidth}|p{0.35\linewidth} ||p{0.14\linewidth}|p{0.35\linewidth}}
\textbf{Attack Label }    & \textbf{Explanation}& \textbf{Safety Label}  & \textbf{Explanation} \\

\hline\hline
\textbf{Server Down}    & Once the attacker has gained access to the server he chooses to deny others access.  & \textbf{Server Down}   & Due to erros with DNS configuration and a faulty update with a failed check the server goes offline\\\hline

\textbf{auth}& Breaching the authentication mechanism & \textbf{faulty update} & Due to unchecked issues in patch the faulty update when't through\\\hline

\textbf{no auth} & Bypassing the authentication mechanism & \textbf{DNS OFF}       & Due to safety procedure DNS hidden from internet \\\hline

\textbf{user privileges} & Gain illegitimate privileges & \textbf{update check}    &  The safety checker for updates failed to spot the fault \\\hline

\textbf{buffer overflow}  & Exploit vulnerabilities to overflow & \textbf{dns check} & Internal safety procedure was that if the internal DNS resolver couldn't find its own address it removed the DNS from the internet \\\hline

\textbf{rsa} & RSAREF2 library buffer overflow vulnerability & \textbf{server patch}  & An engineer send an update to the backbone routers which was faulty \\\hline

\textbf{ssh} & ssh deamon overflow vulnerability & \textbf{resolve dns }  & DNS is unable to resolve its own addresss \\\hline

\textbf{rsh} & Remotely executing shell commands & \textbf{remote login} & Remotely patch the server to remove failure\\\hline
\textbf{ftp} & Exploit ftp vulnerability to upload files&&\\\hline
\textbf{server disrupt}   & Once the attacker has gained access to the server he alters the DNS settings &               & \\\hline
\textbf{disable ssh} & Remove ssh access to stop vulnerability &&\\
\hline             \hline                                                                                                             
\end{tabular}

\end{table}
The triggering of the initial Facebook blackout was caused by two independent events, firstly an engineer sent an update to the backbone routers which was faulty, this was not caught by the safety checker and was allowed to go through.
An internal safety procedure was: that if the internal DNS resolver couldn't find its own address it removed the DNS from the internet, meaning they were invisible to the outside world.
In the case of an attack we consider a generic server, offering ftp, ssh, and rsh services, as seen in previous work~\cite{jhawar2015attack}. 
The attack tree in Fig.~\ref{fig:antagonistic} shows how an attacker can disrupt a server in two ways: either without providing any user credentials (no-auth) or by breaching the authentication mechanism (auth).
In the first case, the attacker must first gain user privileges and then perform a local buffer overflow attack.
To gain user privileges, the attacker must exploit an FTP vulnerability to anonymously upload a list of trusted hosts as well as remotely executing shell commands using RSH. 
The second way is to abuse a buffer overflow in both the ssh daemon (SSH) and the RSAREF2 library (RSA) used for authentication.

Using the described notation we formalise the security and safety scenarios as $DPT_S$ and $DPT_A$ for safety and security respectively:
\begin{equation*}
    DPT_S = ((\text{\texttt{server patch}}\ \hexagon\ \text{\texttt{update check}})\cap(\text{\texttt{resolve DNS}}\ \hexagon\ \text{\texttt{dns check}})) 
\end{equation*}
\begin{equation*}
    DPT_A =  ((\text{\texttt{ftp}}\ \cap\ \text{\texttt{rsh}})\ \cap\ \text{\texttt{buffer overflow}})\ \cup\ (\text{\texttt{rsa}}\ \cap\ \text{\texttt{ssh}})
\end{equation*}

\subsection{Independence}

\begin{figure}[h!]
    \centering
    \includegraphics[width=0.6\textwidth]{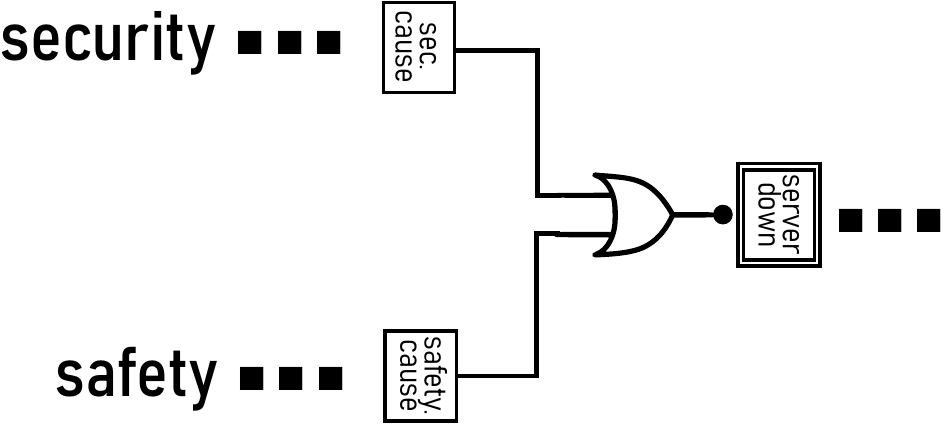}
    \caption{Independence - when the two events do not interact to cause the final outcome. This is a sub-tree of the tree in Fig~\ref{fig:antagonistic}, and as such details are omitted}
    \label{fig:independence}
\end{figure}

The Independence relationship can simply be defined as an \texttt{OR} gate, whereby either the security scenario \textit{or} the safety scenario lead to the eventual risk condition, an initial investigation of this has has been done before~\cite{steiner2013combination}, and can be visualised in Fig.~\ref{fig:independence} for our case study.

For quantitative analysis there are, however, certain complications.
Since now both a safety assessment is present as well as a security assessment, there needs to be a choice as to how the OR gate quantification is performed.
Whilst this topic in itself warrants a deeper discussion orthogonal to the contributions of this work, a naive approach is to employ the approach from Steiner and Liggesmeyer and calculate a composition of risk-probability pairs leading to the \texttt{OR} gate outcome~\cite{steiner2013combination}.
In the case study we simple see that the server outage may be caused by an active attacker (attack tree), or a failure of the update process and DNS misconfiguration (safety tree), or both.

Independence is modelled as two DPTs joining roots with an \texttt{OR} gate.
\begin{definition}
An independent joining of  a security and a safety assessment (assuming both are modelled as disruption defence trees) is defined as: given two disruption defence trees $DPT_{A}$ and $DPT_{F}$ a new tree $DPT_{AF}$ is formed as:
\begin{itemize}
    \item $DPT_{AF} := \langle\, 
        N_A \uplus N_E \uplus \{R_{T_{AF}}\},\ 
        \mathbb{T}, \
        t_a \uplus t_f \uplus \{R_{T_{AF}}\mapsto \text{\texttt{OR}}\},\  
        ch_{A} \uplus ch_{F} \uplus \{R_{T_{AF}} \mapsto [R_{T_{A}}, R_{T_{F}]} \} \rangle$
 \end{itemize}
\label{def:ind}
\end{definition}

So for the example in Fig.~\ref{fig:independence}, the independent joining would look like this:
\begin{equation*}
\begin{split}
     FB_{IND} = &  ( ((\text{\texttt{ftp}}\ \cap\ \text{\texttt{rsa}})\ \cap\ \text{\texttt{buffer overflow}})\ \cup\ (\text{\texttt{rsa}}\ \cap\ \text{\texttt{ssh}}))\ \ \cup \\
 & ((\text{\texttt{server patch}}\ \hexagon\ \text{\texttt{update check}})\cap(\text{\texttt{resolve DNS}}\ \hexagon\ \text{\texttt{dns check}}))
\end{split}
\end{equation*}

\subsection{Conditional Dependency}

\begin{figure} [htbp!]
    \centering
    \includegraphics[width=0.85\textwidth]{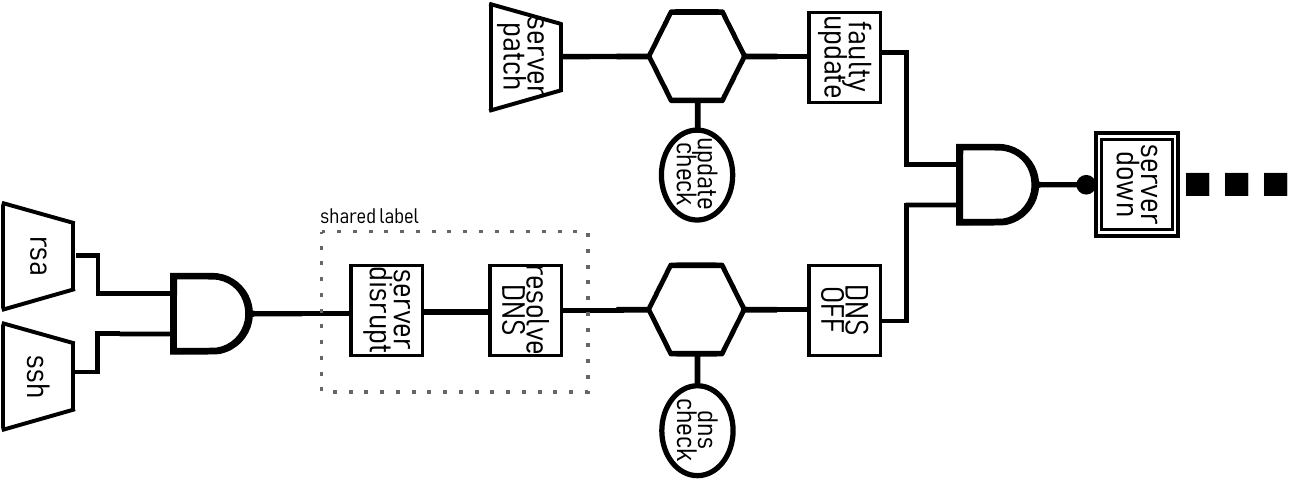}
    \caption{Conditional - when the safety outcome is conditional on the security  or vice versa. Scenario modelled by security  outcomes (simplified) causing fault.}
    \label{fig:conditional}
\end{figure}

Given two disruption prevention trees a conditional dependence means that the vertex of one of the trees corresponds to a child node or intermediate event of the other.
A conditional joining extends the tree to include all the children of the vertex being merged in, with the old vertex simply becoming an intermediate event to the other tree.
The assessment now needs to include all the new nodes that lead to the risk event.
In the example we can easily presume that the failure to resolve the DNS could be due to an attacker having gained access to the server (following the attack path from the attack tree) and changing the configuration.
So by merging the attack tree to the fault tree on the resolve DNS node we get a conditional dependence, as seen in Fig.~\ref{fig:conditional}.

A conditional dependence may be viewed as the expansion of a previously unknown external event.
Whilst previously we had a \texttt{LEAF} node that represented an input to an assessment, we now expand this node into a fully fledged disruption tree.
This notion of external event is seen in previous literature~\cite{vesely1981fault}, and this can be seen as an enhancement of knowledge of what lead to the input events.
Semantically it means that a node at the bottom of a disruption tree simply gains as children all the nodes in another disruption tree, and the previous child leaf node is substituted.

\begin{definition}
A conditional joining of  a security and a safety assessment (assuming both are modelled as a disruption defence tree) is defined as: given a joining of two disruption defence trees $DPT_{A}$ to $DPT_{F}$ and a condition node $n_c$ (of type \INHIBITt) a new tree $DPT_{AF}$ is formed as:
\begin{itemize}
    \item  $DPT_{AF} := \langle N_A \uplus N_F \uplus \{n_c \mapsto R_{T_A}\}, \mathbb{T}, t_A \uplus t_F \uplus \{t_{AF}(n_c) := t_A(R_{T_A})\}, ch_{A} \uplus ch_{F} \rangle$
   
\end{itemize}
\end{definition}

So for the example in Fig.~\ref{fig:conditional}, the conditional joining would look like this:
\begin{equation*}
\begin{split}
     FB_{COND} = & ((\text{\texttt{server patch}}\ \hexagon\ \text{\texttt{update check}})\ \cap \
      ((\text{\texttt{rsa}}\ \cap\ \text{\texttt{ssh}})\ \hexagon\ \text{\texttt{dns check}}))
\end{split}
\end{equation*}

\subsection{Reinforcing Dependency}
\label{sec:reinforce}

\begin{figure} [htbp!]
    \centering
    \includegraphics[width=0.85\textwidth]{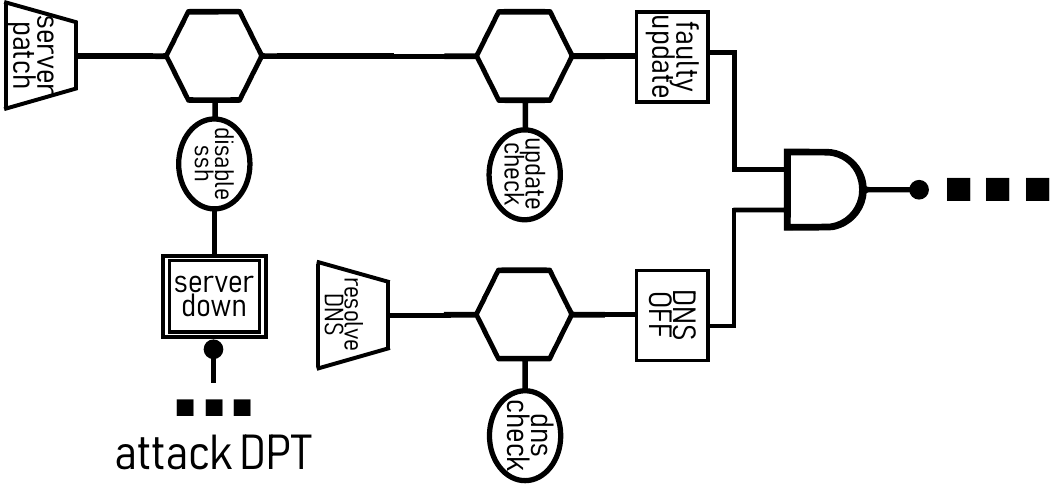}
    \caption{Reinforcing - when a security or safety measure helps the other incident. This is modelled by the outcome of security (simplified) in this case, stopping the safety incident. Attack tree omitted as it is sub-tree of Fig.~\ref{fig:antagonistic}}
    \label{fig:reinforce}
\end{figure}

The reinforcement dependency takes place when a DCT from one bowtie contains a response/outcome that leads to the resolution of the initial scenario of a separate DPT.
This means that due to the response to an incident another incident is no longer plausible.
If we take the same security response as previously, disabling ssh to stop the buffer overflow, we see that the patching of the server event which led to the faulty update cannot take place anymore, as per Fig.~\ref{fig:reinforce}.
This once again requires the right hand side of a bowtie.
In this scenario the response outcome of one bowtie is chaining to the cause event of another tree.
Concretely, the response \texttt{disable ssh} negates the \texttt{server patch} event of the fault tree.
This is modelled through a \texttt{INHIBIT} gate.
Simply this is evaluated as the action no longer being possible (as the response is not negated it results to 0 in \texttt{INHIBIT} semantics).

Reinforcing dependence assumes that as an outcome of fixing a fault or mitigating an attack, another fault or attack is no longer possible.
The approach taken in this scenario is to take the DCT on the right hand side of the bowtie, joining into the reinforced DPT, extending it, and creating a new DPT.
This leads to the final risk event of the reinforced DPT being the new risk event of focus for the whole assessment.
 We make the assumption that if an event does not lead to any other disruptions in the DPT being reinforced it is not influencing the disruption and can be pruned.
 (to preserver the tree structure and avoid back propagating branches). 
We infer that if an event is present in both trees there is an implied relationship, extensions could instead assign further labels to construct these relationships.

\begin{definition}
A reinforcing joining of a disruption consequence assessment (implicitly joining any left most side $DPT$ if a bowtie) $DCT$ and any disruption defence tree $DPT$ takes the form of a joining of the single reinforcing branch $V_C = f_T(R_T \in DCT,C)$ into a node $n_r$ of type \INHIBITt:
\begin{itemize}
    \item  $DPT_{AF} :=  \langle N \uplus V_C  
    , \mathbb{T}
    ,t_P \uplus t_C, %
    ch_P \uplus ch_C \uplus %
    \{ch_{PC}:n_r \mapsto [u1,\overline{V_C}]\}
    $
    
\end{itemize}
\end{definition}

We note that an inherent relationship between prevention and disruptions is assumed. 
This is not automatically inferred through formalisations.
In this case a prevention, child two of the \INHIBITt gate, is replaced by the branch of the DPT that reinforces the first child.
We see that the previous child two $u2$ of the \INHIBITt node is replaced by the branch $V_C$.
One can observe that under this formalism some knowledge about the previous events are lost, as they are pruned in the joining.
For clarity, an example joining between $DCT_i$ and $DPT_j$:
\begin{equation*}
    DPT_i = (\text{\texttt{X}}\ \hexagon\ \text{\texttt{Y}}) \ \text{and}\ DCT_j = ([j_i\cup \overline{j_i},j_n \cup \overline{j_n}])
\end{equation*}
if $j_n$ is reinforcing to \texttt{X}, their reinforcing joining $DPT_{ij}$, joins the path $V_j =\{j_i,j_n\}$ to $DPT_i$
\begin{equation*}
    DPT_{ij} = (\text{\texttt{X}}\ \hexagon\ \overline{V_j} )
\end{equation*}

\subsection{Antagonistic Dependency}

When security and/or safety measures directly antagonise each other it means that a measure that attempts to make the system more secure has an adverse effect on its safety and conversely.
We see that this antagonism is almost exclusively present in an outcome of an incident taking place (i.e. after the warning has gone off), and not visible in the bare DPT.
For this specific scenario we need the use of the right hand side of a bowtie and event trees to represent the dependency.
Going back to our case study, once the server has been taken down, engineers from the safety domain and the security domain investigate the incident.
The first team (safety), sees that the incident is due to a faulty DNS configuration and patches the DNS remotely through SSH.
The second team (security), sees that the incident is due to a vulnerability in the ssh library and therefore disables ssh (perhaps temporarily whilst the issue is being fixed by local engineers, as was the case for Facebook).
As one can see, if the security response is put in place, there is no way the safety response can be put in place, this means that the fault or incident cannot be fixed if the security incident is fixed.
We note this is exactly what took place during the Facebook outage, a security response disallowed the safety engineers to remote in and patch the fault, leading to physical access being required.
In practice this means that the system can either be made safe or it can be made secure, but not both.
This unlike the previous relationships is modelled through the consequence assessment and through the usage of event trees.

\begin{definition}
An antagonistic joining of  a security and a safety consequence is defined as: given a joining of two disruption consequence assessments  $DCT_{A}$ and $DCT_{F}$, and antagonistic event $e$, a new tree $DCT_{AF}$ is formed as:
\begin{itemize}
    \item $DCT_{AF} := \langle N_a \uplus N_s, \mathbb{T},  t_a \uplus t_s, ch_s \uplus ch_a \uplus \{ch_{sa}:e \mapsto \text{\texttt{CHOOSE}} \} \uplus  \{ch_{sa}:\overline{e} \mapsto \text{\texttt{CHOOSE}} \}  \rangle$, 
\end{itemize}
\label{def:antagonistic}
\end{definition}

finally the  Facebook DNS Disruption as a Bowtie can be defined in the following steps, given the $DPT_S$ and $DPT_A$ from Sec.~\ref{sec:example} they can be \textit{independently} joined following Def.~\ref{def:ind} to form:
\begin{equation*}
\begin{split}
 FB_{DPT} = &  ( ((\text{\texttt{ftp}}\ \cap\ \text{\texttt{rsh}})\ \cap\ \text{\texttt{buffer overflow}})\ \cup\ (\text{\texttt{rsa}}\ \cap\ \text{\texttt{ssh}}))\ \ \cup \\
 & ((\text{\texttt{server patch}}\ \hexagon\ \text{\texttt{update check}})\cap(\text{\texttt{resolve DNS}}\ \hexagon\ \text{\texttt{dns check}})) 
 \end{split}
\end{equation*}
\noindent and then given the two possible consequence trees, ${[DCT_A,DCT_S]}$:
\begin{equation*}
   DCT_S = (\text{\texttt{remote login}})\ \text{and}\ DCT_A=(\text{\texttt{disable ssh}})
\end{equation*}
\noindent we can antagonistically join them following Def.~\ref{def:antagonistic} to form the FB case study situation of conflict:
\begin{equation*}
\begin{split}
FB_{DCT} = & (\text{\texttt{remote login}}\ \cap\ \overline{\text{\texttt{disable ssh}}})\ \cup\ (\overline{\text{\texttt{remote login}}}\ \cap\ \text{\texttt{disable ssh}}) \\
& \text{ or more simply }(\text{\texttt{remote login}}\ \oplus\ \text{\texttt{disable ssh}}). 
 \end{split}
\end{equation*}
Putting the two together you can obtain the bowtie $FB_{Bowtie} = [FB_{DPT} \rightarrow [[\text{\texttt{server outage}} ]] \rightarrow FB_{DCT}]$

\section{Conclusions}
\label{sec:conclusions}

This work presented the first formalisation of safety security interdependence covering the full spectrum of dependencies through the use of bowtie diagrams.
We notice some interesting outcomes: 1) for some of the dependencies dynamic gates are necessary i.e. \texttt{INHIBIT} gates (not typically used).
2) some dependencies make more sense on the outcome of the tree, are beyond the attack/fault tree, and need to be reflected in the recovery phase (showcasing the suitability of bowties as a way to explore interdependence).
We also provide the first formalisation of DPTs alongside its semantics.
Finally, this work is the first to provide formalisations of the safety security interdependence operations under a unified assessment.

\subsection{Related work}
This work is by no means the first exploration of combining security and safety~\cite{gould2020safety,kriaa2015survey,kriaa2014safety,abdo2018safety,haider2019concertofla}.
Existing work in the area also highlights that this is by no means an easy feat as there are several differences as an outcome of developing in completely different contexts~\cite{gould2020safety}.
Gould and Bieder~\cite{gould2020safety} explore the initial development of the two fields and the contexts in which they coexist.
Safety is originally developed as means to avoid systems failures guided by the increased understanding of hazards and inevitability of accidents.
Conversely, security developed initially as an outcome of malicious parties wishing to harm systems, however has developed into a critical component of any system analysis.
The authors cite the increase of emphasis on the reduction of risks associated with mitigating security threats as a key factor as to why security and safety should be integrated more.
However security and safety fields diverge vastly on technology usages and quantification of risks.
The two fields however are strongly interconnected and as such their integration is essential.
The authors propose the need for new policies and shared practice as without the other neither field can be complete and neither safe nor secure~\cite{gould2020safety}.
This recent work highlights the need to unify the assessment techniques, although they do not provide intuition as to how to do so.
Literature has already considered how these two fields may be considered under the same methodology~\cite{haider2019concertofla}, however, often this is done rather relying on expert analysis and without underlying formal foundations.
However, there have also been some specific assessment techniques proposed to join the two assessments.

\subsection{Assessment Techniques}

An emerging field that has a deep integration between these two topics is that of industrial control systems (ICS)~\cite{kriaa2015survey}.
Industrial systems have always been focused on safety, however with the new integration of internet infrastructure and the IoT the avenues of attack greatly increase~\cite{arnaboldi2020modelling,arnaboldi2017quantitative}, leading to the need for better security evaluations to be in place.
Work by Kriaa et al.~\cite{kriaa2015survey}, surveys a list of techniques used in this field for both safety and security.
In this work the authors discuss the possibility for a safety case to be broken by security threats and discuss means to address this in the system assessment.
The paper also discusses current standards for safety and security in the context of ICS, how they may integrate and assessments to decide which category needs to be considered.
Using a different formalism of BMDPs previous work has similarly used the case study of a pipeline to observe security and safety interactions~\cite{kriaa2014safety}.
Their work was one of the first to evaluate the different interdependency types. 
In more recent work~\cite{gallina2022multiconcern} address this same topic in the context of safety and security assurance.
In this work the authors conduct an hazard analysis and risk assessment (HARA) a common safety assurance methodology alongside a threat analysis and risk assessment (TARA), which can be used to assess security risks.
Combining these two techniques they are able to conduct a unified safety security assessment for their case study of an autonomous vehicle adhering to the safety standard ISO26262 and security standard ISO21434 for autonomous vehicles.
Although we note that the unification of these techniques is not done formally under a single construct but more as an iterative manual process.

Having established the need for rigourous security assessment recent work experimented with the usage of bowties for this purpose~\cite{bernsmed2017visualizing}.
In their work the authors apply the usage of bowties to security analysis in the maritime industry.
They showcase the flexibility of the approach in quantified various threats and calculating the impact of the attacks. 
The paper proposes a formula for quantifying threat and consequence impact.
The authors also ran a user study showed that people in the maritime industry thought Bowties for security were a very useful tool, showcasing the applicability of bowties to industry case studies.
The first approach combining safety and security bowties~\cite{abdo2018safety}, combines an attack tree with a safety bowtie, and whilst this proves to be a great first step, falls short of completely uniting the various interdependencies into a single assessment.

The literature showcases that there is already a wide use of these techniques across a wide range of fields and a formal representation allows to set the foundations for further expansions and reasoning.
Formal descriptions allow to have a more rigorous way to analyse systems, they allow to establish what you can and can't do using the approach as well as reason about properties of a formalism.
Previous work has applied this in the context of safety cases~\cite{denney2015formal} and showcased it as a valuable way to create and assess system evaluations.
Even more so, security researchers have applied formal representations to attack trees~\cite{jhawar2015attack,mauw2005foundations}, showcasing its usefulness to reason about systems and performing structured analysis~\cite{fila2020exploiting}.
In more recent work~\cite{denney2019role} the authors formalise bowties in and causality sequences for usage of safety and avionics, their formalisation is similar to the one used in this work, although they omit the analysis of interdependency.
As such they do not discuss joining operations and do not assess security and safety under the same construct, which lies at the foundation of our security and safety assessments.
Our work begins to apply the same formal rigour to safety and security assessments using bowties to unify the assessment in a structured and formal manner, both drawing from clear semantic understanding as well as a clear visual representation.

\subsection{Next steps}

In order to fully generalise this approach a larger case study should be adapted, in particular further analysis on how the two types of trees, event and disruption, can be chained to combine the analysis of several consequent risk events is desirable.
We also see as desirable the investigation of further gate types to fully encapsulate the the full spectrum of analysis currently seen across the literature.
The formal definition of joining operations we have provided for security and safety assessments presents a unique potential for automating the process of combining safety security analysis, future work into this area would have tremendous impact to more rigorously examine scenarios such as the Facebook DNS incident from both assessment angles.
However there is also a need for further investigation into properties of these provided join operations.
We also see the usefulness of hierarchicalisation as means to combine sequences of events into a single more readable structure.
This has been explored in the past successfully for assurance cases~\cite{denney2015formal} and as in this approach several assessments are joined together under the same formalism, an enhancement of this approach could be even more effective and desirable. 
We note that further to the formalisation of the structures and their semantics some distinctions exist in how the probabilities of risk are calculated, however this is orthogonal to the work presented here and we reserve this investigation to future work.

\paragraph{Acknowledgements.}  We're grateful to Ewen~Denney for
suggesting to us to investigate bowtie diagrams for safety-security
assessments, as well as comments on an early draft.
This work was funded by the AISEC grant under EPSRC number EP/T027037/1
 
\bibliographystyle{splncs04}
\bibliography{bib}

\begin{thebibliography}{10}
\providecommand{\url}[1]{\texttt{#1}}
\providecommand{\urlprefix}{URL }
\providecommand{\doi}[1]{https://doi.org/#1}

\bibitem{abdo2018safety}
Abdo, H., Kaouk, M., Flaus, J.M., Masse, F.: A safety/security risk analysis
  approach of industrial control systems: A cyber bowtie--combining new version
  of attack tree with bowtie analysis. Computers \& security  \textbf{72},
  175--195 (2018)

\bibitem{arnaboldi2020modelling}
Arnaboldi, L., Czekster, R.M., Morisset, C., Metere, R.: Modelling
  load-changing attacks in cyber-physical systems. Electronic Notes in
  Theoretical Computer Science  \textbf{353},  39--60 (2020)

\bibitem{arnaboldi2017quantitative}
Arnaboldi, L., Morisset, C.: Quantitative analysis of dos attacks and client
  puzzles in iot systems. In: International Workshop on Security and Trust
  Management. pp. 224--233. Springer (2017)

\bibitem{bernsmed2017visualizing}
Bernsmed, K., Fr{\o}ystad, C., Meland, P.H., Nesheim, D.A., R{\o}dseth,
  {\O}.J.: Visualizing cyber security risks with bow-tie diagrams. In:
  International Workshop on Graphical Models for Security. pp. 38--56. Springer
  (2017)

\bibitem{budde2021attack}
Budde, C.E., Kolb, C., Stoelinga, M.: Attack trees vs. fault trees: two sides
  of the same coin from different currencies. In: International Conference on
  Quantitative Evaluation of Systems. pp. 457--467. Springer (2021)

\bibitem{de2006aramis}
De~Dianous, V., Fievez, C.: Aramis project: A more explicit demonstration of
  risk control through the use of bow--tie diagrams and the evaluation of
  safety barrier performance. Journal of Hazardous Materials  \textbf{130}(3),
  220--233 (2006)

\bibitem{denney2015formal}
Denney, E., Pai, G., Whiteside, I.: Formal foundations for hierarchical safety
  cases. In: 2015 IEEE 16th International Symposium on High Assurance Systems
  Engineering. pp. 52--59. IEEE (2015)

\bibitem{denney2019role}
Denney, E., Pai, G., Whiteside, I.: The role of safety architectures in
  aviation safety cases. Reliability Engineering \& System Safety
  \textbf{191},  106502 (2019)

\bibitem{eames1999integration}
Eames, D.P., Moffett, J.: The integration of safety and security requirements.
  In: International Conference on Computer Safety, Reliability, and Security.
  pp. 468--480. Springer (1999)

\bibitem{fila2020exploiting}
Fila, B., Wide{\l}, W.: Exploiting attack--defense trees to find an optimal set
  of countermeasures. In: 2020 IEEE 33rd Computer Security Foundations
  Symposium (CSF). pp. 395--410. IEEE (2020)

\bibitem{gallina2022multiconcern}
Gallina, B., Montecchi, L., de~Oliveira, A.L., Bressan, L.P.: Multiconcern
  dependability-centered assurance via qualitative and quantitative coanalysis.
  IEEE Software  (2022)

\bibitem{gould2020safety}
Gould, K.P., Bieder, C.: Safety and security: The challenges of bringing them
  together. In: The Coupling of Safety and Security, pp.~1--8. Springer, Cham
  (2020)

\bibitem{guldenmund2006development}
Guldenmund, F., Hale, A., Goossens, L., Betten, J., Duijm, N.J.: The
  development of an audit technique to assess the quality of safety barrier
  management. Journal of hazardous materials  \textbf{130}(3),  234--241 (2006)

\bibitem{haider2019concertofla}
Haider, Z., Gallina, B., Carlsson, A., Mazzini, S., Puri, S.: Concertofla-based
  multi-concern assurance for space systems. ADA USER  \textbf{40}(1), ~35
  (2019)

\bibitem{janardhan2021first}
Janardhan, S.: Update about the october 4th outage (Oct 2021),
  \url{https://engineering.fb.com/2021/10/04/networking-traffic/outage/}

\bibitem{janardhan2021second}
Janardhan, S., Janardhan, S.: More details about the october 4 outage (Oct
  2021),
  \url{https://engineering.fb.com/2021/10/05/networking-traffic/outage-details/}

\bibitem{jhawar2015attack}
Jhawar, R., Kordy, B., Mauw, S., Radomirovi{\'c}, S., Trujillo-Rasua, R.:
  Attack trees with sequential conjunction. In: IFIP International Information
  Security and Privacy Conference. pp. 339--353. Springer (2015)

\bibitem{kordy2010foundations}
Kordy, B., Mauw, S., Radomirovi{\'c}, S., Schweitzer, P.: Foundations of
  attack--defense trees. In: International Workshop on Formal Aspects in
  Security and Trust. pp. 80--95. Springer (2010)

\bibitem{kriaa2014safety}
Kriaa, S., Bouissou, M., Colin, F., Halgand, Y., Pietre-Cambacedes, L.: Safety
  and security interactions modeling using the bdmp formalism: case study of a
  pipeline. In: International Conference on Computer Safety, Reliability, and
  Security. pp. 326--341. Springer (2014)

\bibitem{kriaa2015survey}
Kriaa, S., Pietre-Cambacedes, L., Bouissou, M., Halgand, Y.: A survey of
  approaches combining safety and security for industrial control systems.
  Reliability engineering \& system safety  \textbf{139},  156--178 (2015)

\bibitem{martinho2021}
Martinho, C.: Understanding how {Facebook} disappeared from the {Internet} (Oct
  2021), \url{https://blog.cloudflare.com/october-2021-facebook-outage/}

\bibitem{mauw2005foundations}
Mauw, S., Oostdijk, M.: Foundations of attack trees. In: International
  Conference on Information Security and Cryptology. pp. 186--198. Springer
  (2005)

\bibitem{nielsen1971cause}
Nielsen, D.S.: The cause/consequence diagram method as a basis for quantitative
  accident analysis. Ris{\o} National Laboratory (1971)

\bibitem{ren2017fault}
Ren, H., Chen, X., Chen, Y.: Fault tree analysis for composite structural
  damage. In: Reliability Based Aircraft Maintenance Optimization Applications,
  pp. 115--131. Academic (2017)

\bibitem{de2016bowtie}
de~Ruijter, A., Guldenmund, F.: The bowtie method: A review. Safety science
  \textbf{88},  211--218 (2016)

\bibitem{schneier1999attack}
Schneier, B.: Attack trees. Dr. Dobb’s journal  \textbf{24}(12),  21--29
  (1999)

\bibitem{shahriar2012risk}
Shahriar, A., Sadiq, R., Tesfamariam, S.: Risk analysis for oil \& gas
  pipelines: A sustainability assessment approach using fuzzy based bow-tie
  analysis. Journal of loss prevention in the process Industries
  \textbf{25}(3),  505--523 (2012)

\bibitem{steiner2013combination}
Steiner, M., Liggesmeyer, P.: Combination of safety and security
  analysis-finding security problems that threaten the safety of a system. In:
  DECS : ERCIM/EWICS Workshop on Dependable Embedded and Cyber-physical Systems
  (2013)

\bibitem{stoelinga2021marriage}
Stoelinga, M., Kolb, C., Nicoletti, S.M., Budde, C.E., Hahn, E.M.: The marriage
  between safety and cybersecurity: Still practicing. In: International
  Symposium on Model Checking Software. pp. 3--21. Springer (2021)

\bibitem{vesely1981fault}
Vesely, W.E., Goldberg, F.F., Roberts, N.H., Haasl, D.F.: Fault tree handbook.
  Tech. Rep. NUREG-0492, Nuclear Regulatory Commission Washington DC (1981)

\end{thebibliography}
\end{document}